\documentclass[a4paper,10pt,twoside]{cpc-hepnp}

\usepackage{multicol}
\usepackage{graphicx}
\usepackage{booktabs}
\usepackage{amssymb,bm,mathrsfs,bbm,amscd}
\usepackage[tbtags]{amsmath}
\usepackage{lastpage}

\newcommand{ \be}{\begin{equation}}
\newcommand{ \ee}{\end{equation}}
\newcommand{ \bea}{\begin{eqnarray}}
\newcommand{ \eea}{\end{eqnarray}}
\newcommand{ \mysmall}[1]{\scriptscriptstyle #1} 
\newcommand{ \amu}{a_{\mu}}
\newcommand{ \mw}{M_{\mysmall{W}}}
\newcommand{ \mz}{M_{\mysmall{Z}}}
\newcommand{ \mh}{M_{\mysmall{H}}}
\newcommand{ \mhUB}{M_{\mysmall{H}}^{\mysmall \rm UB}}
\newcommand{ \mhLB}{M_{\mysmall{H}}^{\mysmall \rm LB}}
\newcommand{ \mt}{M_{t}}
\newcommand{ \mpi}{m_{\pi}}
\newcommand{ \seff}{\sin^2 \!\theta_{\rm eff}^{\rm lept}}
\newcommand{ \eq}[1]{Eq.~(\ref{eq:#1})}

\newcommand{ \bom}   {\boldmath}
\newcommand{ \ubm}  {\unboldmath}
\newcommand{ \gmt}  {$g$$-$$2$~}
\newcommand{ \dafive}{\Delta \alpha_{\rm had}^{\mysmall{(5)}}(\mz)}

\begin{document}

\fancyhead[co]{\footnotesize M.~Passera, W.J.~Marciano, A.~Sirlin: The muon
  $g-2$ discrepancy: new physics or a relatively light
  Higgs?}

\footnotetext[0]{Received January $25^{\rm th}$ 2010}

\title{The muon \bom $g-2$ \ubm discrepancy: new physics\\
  or a relatively light Higgs?}

\author{%
        M.~Passera$^{1}$%
\quad W.~J.~Marciano$^{2}$%
\quad A.~Sirlin$^{3}$%
}
\maketitle

\address{%
  ${^1}$~{\it Istituto Nazionale Fisica Nucleare,
  Sezione di Padova, I-35131, Padova, Italy}\\
  ${^2}$~{\it Brookhaven National Laboratory,
  Upton, New York 11973, USA}\\
  ${^3}$~{\it Department of Physics, New York University, 10003
  New York NY, USA}\\
 }

\begin{abstract}
  After a brief review of the muon \gmt status, we discuss
  hypothetical errors in the Standard Model prediction that
  might explain the present discrepancy with the
  experimental value. None of them seems likely. In
  particular, a hypothetical increase of the hadroproduction
  cross section in low-energy $e^+e^-$ collisions could
  bridge the muon \gmt discrepancy, but it is shown to be
  unlikely in view of current experimental error
  estimates. If, nonetheless, this turns out to be the
  explanation of the discrepancy, then the 95\% CL upper
  bound on the Higgs boson mass is reduced to about 135~GeV
  which, in conjunction with the experimental 114.4~GeV 95\%
  CL lower bound, leaves a narrow window for the mass of
  this fundamental particle.
\end{abstract}

\begin{keyword}
  Muon anomalous magnetic moment, Standard Model Higgs boson
\end{keyword}

\begin{pacs}
  13.40.Em, 14.60.Ef, 12.15.Lk, 14.80.Bn
\end{pacs}

\begin{multicols}{2}

\section{Introduction: status of \bom $\amu$ \ubm}

The anomalous magnetic moment of the muon, $a_{\mu}$, is one
of the most interesting observables in particle physics.
Indeed, as each sector of the Standard Model ({\small SM})
contributes in a significant way to its theoretical
prediction, the precise $a_{\mu}$ measurement by the E821
experiment at Brookhaven~\cite{bnl,RobertsPHIPSI09} allows us to test the
entire {\small SM} and scrutinize viable ``new physics''
appendages to this theory~\cite{Czarnecki:2001pv,Stockinger:2006zn}. 

The {\small SM} prediction of the muon \gmt is conveniently
split into {\small QED}, electroweak ({\small EW}) and
hadronic (leading- and higher-order) contributions:
$
    a_{\mu}^{\mysmall \rm SM} = 
         a_{\mu}^{\mysmall \rm QED} +
         a_{\mu}^{\mysmall \rm EW}  +
         a_{\mu}^{\mbox{$\scriptscriptstyle{\rm HLO}$}} +
         a_{\mu}^{\mbox{$\scriptscriptstyle{\rm HHO}$}}.
$  
The {\small QED} prediction, computed up to four (and
estimated at five) loops, currently stands at
$a_{\mu}^{\mysmall \rm QED} = 116584718.08(15)
\times 10^{-11}$\cite{QED}, 
while the {\small EW} effects provide
$a_{\mu}^{\mysmall \rm EW} = 154(2) \times 10^{-11}$\cite{EW}.
The latest calculations of the hadronic leading-order
contribution, via the hadronic $e^+ e^-$ annihilation data,
are in agreement:
$ a_{\mu}^{\mbox{$\scriptscriptstyle{\rm HLO}$}} = 
6894(40) \times 10^{-11}$\cite{TeubnerPHIPSI09} 
(this preliminary result, presented at this workshop,
updates the value $6894(46) \times 10^{-11}$ of
Ref.~\cite{HMNT06}) and
$6903(53) \times 10^{-11}$\cite{Jegerlehner:2009ry}.
These determinations include the 2008 $e^+e^- \to \pi^+
\pi^- (\gamma)$ cross section data from {\small
  KLOE}~\cite{KLOE08} (see also~\cite{MuellerPHIPSI09}). A
somewhat larger value,
$6955(41) \times 10^{-11}$\cite{Davier:2009zi},
was recently obtained including also the 2009 $\pi^+ \pi^-
(\gamma)$ data of BaBar~\cite{Babar2pi}.

The higher-order hadronic term is further divided into two
parts:
$
     \amu^{\mbox{$\scriptscriptstyle{\rm HHO}$}}=
     \amu^{\mbox{$\scriptscriptstyle{\rm HHO}$}}(\mbox{vp})+
     \amu^{\mbox{$\scriptscriptstyle{\rm HHO}$}}(\mbox{lbl}).
$
The first one, 
$-98\, (1) \times 10^{-11}$\cite{HMNT06},
is the $O(\alpha^3)$ contribution of diagrams containing
hadronic vacuum polarization insertions~\cite{Kr96}. The
second term, also of $O(\alpha^3)$, is the hadronic
light-by-light contribution; as it cannot be determined from
data, its evaluation relies on specific models. The latest
determinations of this term,
$116(39) \times 10^{-11}$\cite{Jegerlehner:2009ry, Nyffeler:2009tw}
and
$105(26) \times 10^{-11}$\cite{Prades:2009tw},
are in very good agreement.  If we add the latter to
$a_{\mu}^{\mbox{$\scriptscriptstyle{\rm HLO}$}}$, for
example the value of Ref.~\cite{TeubnerPHIPSI09}, and the
rest of the {\small SM} contributions, we obtain
$  \amu^{\mbox{$\scriptscriptstyle{\rm SM}$}}= 116591773(48)  
\times 10^{-11}$.
The difference with the experimental value
$
    a_{\mu}^{\mbox{$\scriptscriptstyle{\rm EXP}$}}  =
               116592089(63) \times 10^{-11}
$~\cite{RobertsPHIPSI09}
(note the tiny shift upwards, with respect to the value
reported in~\cite{bnl}, due to the updated value of the
muon-proton magnetic moment ratio~\cite{Mohr:2008fa}) is
$\Delta a_{\mu} = a_{\mu}^{\mbox{$\scriptscriptstyle{\rm EXP}$}}-
\amu^{\mbox{$\scriptscriptstyle{\rm SM}$}} = +316(79) \times 10^{-11}$,
i.e., 4.0$\sigma$ (all errors were added in quadrature).
Slightly smaller discrepancies are found employing the
$a_{\mu}^{\mbox{$\scriptscriptstyle{\rm HLO}$}}$ values
reported in~\cite{Davier:2009zi} (which also includes the
recent $\pi^+ \pi^- (\gamma)$ data of BaBar) and
\cite{Jegerlehner:2009ry}: 3.2$\sigma$ and 3.6$\sigma$,
respectively.  We will use the
$a_{\mu}^{\mbox{$\scriptscriptstyle{\rm HLO}$}}$ value of
Ref.~\cite{TeubnerPHIPSI09} (which also provides the
hadronic contribution to the effective fine-structure
constant later required for our analysis), but we expect
that a consistent inclusion of the recent $\pi^+ \pi^-
(\gamma)$ BaBar data would not change our basic
conclusions. For reviews of $\amu$ see
Refs.~\cite{TeubnerPHIPSI09,Jegerlehner:2009ry,Reviews}.

The term $a_{\mu}^{\mbox{$\scriptscriptstyle{\rm HLO}$}}$
can alternatively be computed incorporating hadronic
$\tau$-decay data, related to those of hadroproduction in
$e^+e^-$ collisions via isospin symmetry~\cite{ADH98}.  The
long-standing difference between the $e^+e^-$- and
$\tau$-based determinations of
$a_{\mu}^{\mbox{$\scriptscriptstyle{\rm HLO}$}}$~\cite{DEHZ}
has been recently somewhat lessened by a
re-analysis~\cite{Davier:2009ag} where the isospin-breaking
corrections~\cite{IVC} were revisited taking advantage of
more accurate data and new theoretical investigations
(recent $\tau^- \to \pi^- \pi^0 \nu_{\tau}$ data from the
Belle experiment~\cite{Fujikawa:2008ma} were also
included). In spite of this, the $\tau$-based value remains
higher than the $e^+e^-$-based one, leading to a smaller
($1.9 \sigma$) difference $\Delta a_{\mu}$. On the other
hand, recent analyses of the pion form factor claim that the
$\tau$ and $e^+e^-$ data are consistent after isospin
violation effects and vector meson mixings are considered,
further confirming the $e^+e^-$-based
discrepancy~\cite{Benayoun}.

The 3--4$\sigma$ discrepancy between the theoretical
prediction and the experimental value of the muon \gmt can
be explained in several ways. It could be due, at least in
part, to an error in the determination of the hadronic
light-by-light contribution. However, if this were the only
cause of the discrepancy, $a_{\mu}^{\mysmall \rm
  HHO}(\mbox{lbl})$ would have to move up by many standard
deviations (roughly ten) to fix it.  Although the errors
assigned to $a_{\mu}^{\mysmall \rm HHO}(\mbox{lbl})$ are
only educated guesses, this solution seems unlikely, at
least as the dominant one.

Another possibility is to explain the discrepancy $\Delta
a_{\mu}$ via the {\small QED}, {\small EW} and hadronic
higher-order vacuum polarization contributions; this looks
very improbable, as one can immediately conclude inspecting
their values and uncertainties reported above. If we assume
that the \gmt experiment {\small E821} is correct, we are
left with two options: possible contributions of physics
beyond the {\small SM}, or an erroneous determination of the
leading-order hadronic contribution
$a_{\mu}^{\mbox{$\scriptscriptstyle{\rm HLO}$}}$ (or
both). The first of these two explanations has been
extensively discussed in the literature; updating
Ref.~\cite{Passera:2008jk} we will study whether the second
one is realistic or not, and analyze its implications for
the {\small EW} bounds on the mass of the Higgs boson.

\section{Connection with the Higgs mass}

The hadronic leading-order contribution
$\amu^{\mbox{$\scriptscriptstyle{\rm HLO}$}}$ can be
computed via the dispersion integral~\cite{DISPamu}
\be
      a_{\mu}^{\mbox{$\scriptscriptstyle{\rm HLO}$}}= 
      \frac{1}{4\pi^3} \!
      \int^{\infty}_{4m_{\pi}^2} ds \, K(s) \, \sigma (s),
\label{eq:amudispint}
\ee
where $\sigma (s)$ is the total cross section for $e^+ e^-$
annihilation into any hadronic state, with vacuum
polarization and initial state {\small QED} corrections
subtracted off (for a detailed discussion of these radiative
corrections and the precision of the Monte Carlo generators
used to analyze the hadronic cross section measurements
see~\cite{Actis:2009gg}), and $s$ is the squared momentum
transfer. The well-known kernel function $K(s)$
(see~\cite{EJ95}) is positive definite, decreases
monotonically for increasing $s$ and, for large $s$, behaves
as $m_\mu^2/(3s)$ to a good approximation. About 90\% of the
total contribution to $\amu^{\mbox{$\scriptscriptstyle{\rm
      HLO}$}}$ is accumulated at center-of-mass energies
$\sqrt{s}$ below 1.8~GeV and roughly three-fourths of
$\amu^{\mbox{$\scriptscriptstyle{\rm HLO}$}}$ is covered by
the two-pion final state which is dominated by the
$\rho(770)$ resonance~\cite{Davier:2009zi}. Exclusive
low-energy $e^+e^-$ cross sections were measured at
colliders in Frascati, Novosibirsk, Orsay, and Stanford,
while at higher energies the total cross section was
determined inclusively.

Let's now assume that the discrepancy
$\Delta a_{\mu} = a_{\mu}^{\mbox{$\scriptscriptstyle{\rm EXP}$}}-
\amu^{\mbox{$\scriptscriptstyle{\rm SM}$}} = +316(79) \times 10^{-11}$,
is due to --~and only to~-- hypothetical errors in $\sigma
(s)$, and let us increase this cross section in order to
raise $\amu^{\mbox{$\scriptscriptstyle{\rm HLO}$}}$, thus
reducing $\Delta a_{\mu}$. This simple assumption leads to
interesting consequences. An upward shift of the hadronic
cross section also induces an increase of the value of the
hadronic contribution to the effective fine-structure
constant at $M_Z$~\cite{DISPDalpha},
\be
 \dafive = \frac{\mz^2}{4 \alpha \pi^2}
  \,\, P \! \int_{4m_\pi^2}^{\infty} ds \, \frac{\sigma(s)}{\mz^2 -s}
\label{eq:Dpi5dispint}
\ee       
($P$ stands for Cauchy's principal value).  This integral is
similar to the one we encountered in \eq{amudispint} for
$a_{\mu}^{\mbox{$\scriptscriptstyle{\rm HLO}$}}$. There,
however, the weight function in the integrand gives a
stronger weight to low-energy data.
Let us define
\be
     a_i = \int_{4m_{\pi}^2}^{s_u}ds \, f_i(s) \, \sigma (s)
\ee
$(i=1,2)$, where the upper limit of integration is $s_u <
\mz^2$, and the kernels are $f_1(s) = K(s)/(4 \pi^3)$ and
$f_2(s) = [\mz^2/(\mz^2-s)]/(4 \alpha \pi^2)$. The integrals
$a_i$ with $i=1,2$ provide the contributions to
$a_{\mu}^{\mbox{$\scriptscriptstyle{\rm HLO}$}}$ and
$\dafive$, respectively, from $4m_\pi^2$ up to $s_u$ (see
Eqs.~(\ref{eq:amudispint},\ref{eq:Dpi5dispint})).
An increase of the cross section $\sigma(s)$ of the form
\be
     \Delta \sigma(s) = \epsilon \sigma(s)
\ee
in the energy range $\sqrt s \in [\sqrt s_0 - \delta/2,
\sqrt s_0 + \delta/2]$, where $\epsilon$ and $\delta$ are positive
constants and $2m_{\pi}+\delta/2<\sqrt s_0<\sqrt s_u
-\delta/2$, increases $a_1$ by 
$\Delta a_1 (\sqrt s_0,\delta,\epsilon) = \epsilon
\int_{\sqrt s_0-\delta/2}^{\sqrt s_0+\delta/2} 2t \,
\sigma(t^2) \, f_1(t^2) \, dt$.
If we assume that the muon \gmt discrepancy is entirely due
to this increase in $\sigma(s)$, so that $\Delta a_1 (\sqrt
s_0,\delta,\epsilon) = \Delta a_{\mu}$, the parameter
$\epsilon$ becomes
\be
       \epsilon =  
       \frac{\Delta a_{\mu}}{
         \int_{\sqrt s_0-\delta/2}^{\sqrt s_0 +\delta/2} 
         2t \, f_1(t^2) \, \sigma(t^2) \, dt},
\label{eq:eps} 
\ee
and the corresponding increase in $\dafive$ is
\be 
\Delta a_2(\sqrt s_0,\delta) = \Delta a_{\mu}
\frac{\int_{\sqrt s_0-\delta/2}^{\sqrt s_0+\delta/2}
  f_2(t^2) \, \sigma(t^2) \, t \, dt} {\int_{\sqrt
    s_0-\delta/2}^{\sqrt s_0+\delta/2} f_1(t^2) \,
  \sigma(t^2) \, t \, dt}.
\label{eq:shiftb} 
\ee
The shifts $\Delta a_2(\sqrt s_0,\delta)$ were studied in
Ref.~\cite{Passera:2008jk} for several bin widths $\delta$
and central values $\sqrt s_0$.

The present global fit of the {\small LEP} Electroweak
Working Group ({\small EWWG}) leads to the Higgs boson mass
$\mh \!=\! 87^{+35}_{-26}$~GeV 
and the 95\% confidence level ({\small CL}) upper bound
$\mhUB \!\simeq\! 157$~GeV~\cite{newLEPEWWG}. This result is
based on the recent preliminary top quark mass
$\mt\!=\!173.1(1.3)$~GeV~\cite{Tevatron2009} and the value
$\dafive \!=\! 0.02758(35)$~\cite{BP05}.
The {\small LEP} direct-search 95\%{\small CL} lower bound
is $\mhLB=114.4$~GeV~\cite{MHLB03}.
Although the global {\small EW} fit employs a large set of
observables, $\mhUB$ is strongly driven by the comparison of
the theoretical predictions of the W boson mass and the
effective {\small EW} mixing angle $\seff$ with their
precisely measured values. Convenient formulae providing the
$\mw$ and $\seff$ {\small SM} predictions in terms of $\mh$,
$\mt$, $\dafive$, and $\alpha_s(\mz)$, the strong coupling
constant at the scale $\mz$, are given in~\cite{formulette}.
Combining these two predictions via a numerical
$\chi^2$-analysis and using the present world-average values
$\mw \!=\! 80.399(23)$~GeV~\cite{Wmass}, 
$\seff \!=\! 0.23153(16)$~\cite{LEPEWWG05},
$\mt \!=\! 173.1(1.3)$~GeV~\cite{Tevatron2009}, 
$\alpha_s(\mz) \!=\! 0.118(2)$~\cite{PDG08},
and the determination
$\dafive = 0.02758(35)$~\cite{BP05},
we get
$\mh = 92^{+37}_{-28}$~GeV
and $\mhUB=158$~GeV. We see that indeed the $\mh$ values
obtained from the $\mw$ and $\seff$ predictions are quite
close to the results of the global analysis.

The $\mh$ dependence of $\amu^{\mbox{$\scriptscriptstyle{\rm
      SM}$}}$ is too weak to provide $\mh$ bounds from the
comparison with the measured value.
On the other hand, $\dafive$ is one of the key inputs of the
{\small EW} fits. For example, employing the latest
(preliminary) value
$\dafive = 0.02760(15)$
presented at this workshop~\cite{TeubnerPHIPSI09} instead of
0.02758(35)~\cite{BP05}, the $\mh$ prediction derived from
$\mw$ and $\seff$ shifts to
$\mh = 96^{+32}_{-25}$~GeV 
and $\mhUB=153$~GeV. To update the analysis of
Ref.~\cite{Passera:2008jk} we considered the new values of
$\dafive$ obtained shifting
0.02760(15)~\cite{TeubnerPHIPSI09} by $\Delta a_2(\sqrt s_0,
\delta)$ (including their uncertainties, as discussed
in~\cite{Passera:2008jk}), and computed the corresponding
new values of $\mhUB$ via the combined $\chi^2$-analysis
based on the $\mw$ and $\seff$ inputs (for both $\dafive$
and $\amu^{\mbox{$\scriptscriptstyle{\rm HLO}$}}$ we used
the values reported in~\cite{TeubnerPHIPSI09}). Our results
show that an increase
$\epsilon \sigma (s)$
of the hadronic cross section (in $\sqrt s \in [\sqrt s_0 -
\delta/2, \sqrt s_0 + \delta/2]$), adjusted to bridge the
muon \gmt discrepancy $\Delta a_{\mu}$, decreases $\mhUB$,
further restricting the already narrow allowed region for
$\mh$. We conclude that these hypothetical shifts conflict
with the lower limit $\mhLB$ when
$\sqrt s_0 \gtrsim 1.2$~GeV, 
for values of $\delta$ up to several hundreds of
MeV. In~\cite{Passera:2008jk} we pointed out that there are
more complex scenarios where it is possible to bridge the
$\Delta a_{\mu}$ discrepancy without significantly affecting
$\mhUB$, but they are considerably more unlikely than those
discussed above.

If $\tau$ data are used instead of $e^+ e^-$ ones in the
calculation of the dispersive integral in
Eq.~(\ref{eq:amudispint}),
$a_{\mu}^{\mbox{$\scriptscriptstyle{\rm HLO}$}}$ increases
to
$7053(45) \times 10^{-11}$\cite{Davier:2009ag}
and the discrepancy drops to
$\Delta a_{\mu} = +157(82) \times 10^{-11}$,
i.e.\ $1.9\sigma$. While using $\tau$ data reduces the
$\Delta a_{\mu}$ discrepancy, it increases $\dafive$ by
approximately $2 \times 10^{-4}$,\footnote{~This number
  represents our rough update of the value reported in
  Ref.~\cite{DEHZ}.} leading to a sharply lower $\mh$
prediction~\cite{Marciano04}. Indeed, increasing the
previously employed value
$\dafive =0.02760(15)$~\cite{TeubnerPHIPSI09} 
by $2 \times 10^{-4}$ and using the same above-discussed
previous inputs of the $\chi^2$-analysis, we find an $\mhUB$
value of only 138~GeV. If the remaining 1.9$\sigma$
discrepancy $\Delta a_{\mu}$ is bridged by a further
increase $\Delta \sigma(s) = \epsilon \sigma(s)$ of the
hadronic cross section, $\mhUB$ decreases to even lower
values, leading to a scenario in near conflict with $\mhLB$.

Recent analyses of the pion form factor below 1~GeV claim
that $\tau$ data are consistent with the $e^+e^-$ ones after
isospin violation effects and vector meson mixings are
considered~\cite{Benayoun}. In this case one could use the
$e^+e^-$ data below $\sim \!1$~GeV, confirmed by the $\tau$
ones, and assume that $\Delta a_{\mu}$ is accommodated by
hypothetical errors in the $e^+e^-$ measurements occurring
above $\sim \!1$~GeV, where disagreement persists between
these two data sets. Our analysis shows that this assumption
would lead to $\mhUB$ values inconsistent with $\mhLB$.

In the above analysis, the hadronic cross section
$\sigma(s)$ was shifted up by amounts $\Delta \sigma(s) =
\epsilon \sigma(s)$ adjusted to bridge $\Delta
a_{\mu}$. Apart from the implications for $\mh$, these
shifts may actually be inadmissibly large when compared with
the quoted experimental uncertainties. Consider the
parameter $\epsilon=\Delta \sigma(s)/\sigma(s)$. Clearly,
its value depends on the choice of the energy range $[\sqrt
s_0 - \delta/2, \sqrt s_0 + \delta/2]$ where $\sigma(s)$ is
increased and, for fixed $\sqrt s_0$, it decreases when
$\delta$ increases. Its minimum value, $\sim 5\%$, occurs if
$\sigma(s)$ is multiplied by $(1+\epsilon)$ in the whole
integration region, from $2\mpi$ to infinity. Such a shift
would lead to $\mhUB \sim 75$~GeV, well below $\mhLB$.
Higher values of $\epsilon$ are obtained for narrower energy
bins, particularly if they do not include the
$\rho$-$\omega$ resonance region. For example, a huge
$\epsilon \sim 55\%$ increase is needed to accommodate
$\Delta a_{\mu}$ with a shift of $\sigma(s)$ in the region
from $2\mpi$ up to 500~MeV (reducing $\mhUB$ to 146~GeV),
while an increase in a bin of the same size but centered at
the $\rho$ peak requires $\epsilon \sim 9\%$ (lowering
$\mhUB$ to 135~GeV). As the quoted experimental uncertainty
of $\sigma(s)$ below 1~GeV is of the order of a few per cent
(or less, in some specific energy regions), the possibility
to explain $\Delta a_{\mu}$ with these shifts $\Delta
\sigma(s)$ appears to be unlikely. Lower values of
$\epsilon$ are obtained if the shifts occur in energy ranges
centered around the $\rho$-$\omega$ resonances, but also
this possibility looks unlikely, since it requires
variations of $\sigma(s)$ of at least $\sim 6$\%. If,
however, such shifts $\Delta \sigma(s)$ indeed turn out to
be the solution of the $\Delta a_{\mu}$ discrepancy, then
$\mhUB$ is reduced to about 135~GeV.

It is interesting to note that in the scenario where $\Delta
a_{\mu}$ is due to hypothetical errors in $\sigma(s)$,
rather than ``new physics'', the reduced $\mhUB \lesssim
135$~GeV induces some tension with the approximate 95\%
{\small CL} lower bound $\mh \gtrsim 120$~GeV required to
ensure vacuum stability under the assumption that the
{\small SM} is valid up to the Planck
scale~\cite{VacuumStability} (note, however, that this lower
bound somewhat decreases when the vacuum is allowed to be
metastable, provided its lifetime is longer than the age of
the universe~\cite{VacuumMetaStability}). Thus, one could
argue that this tension is, on its own, suggestive of
physics beyond the {\small SM}.

We remind the reader that the present values of $\seff$
derived from the leptonic and hadronic observables are
respectively $(\seff)_{l} \! = \! 0.23113(21)$ and
$(\seff)_{h} \! = \! 0.23222(27)$~\cite{LEPEWWG05}. In
Ref.~\cite{Passera:2008jk} we pointed out that the use of
either of these values as an input parameter leads to
inconsistencies in the {\small SM} framework that already
require the presence of ``new physics''. For this reason, we
followed the standard practice of employing as input the
world-average value for $\seff$ determined in the {\small
  SM} global analysis. Since $\mhUB$ also depends
sensitively on $\mt$, in~\cite{Passera:2008jk} we provided
simple formulae to obtain the new values derived from
different $\mt$ inputs.

A 3--4$\sigma$ discrepancy between the theoretical
prediction and the experimental value of the muon \gmt would
have interesting implications if truly due to ``new
physics'' (i.e.\ beyond the {\small SM} expectations).
Supersymmetry provides a natural interpretation of this
discrepancy (see Ref.~\cite{Stockinger:2006zn} for a
review). For illustration purposes, we assume a single mass
$m_{\scriptstyle \rm susy}$ for sleptons, sneutrinos and
gauginos that enter the $\amu^{\scriptstyle \rm susy}$
calculation. Then one finds~\cite{susy} (including leading
two-loop effects)
\be
       a_{\mu}^{\scriptstyle \rm susy} \simeq 
       \mbox{\rm sgn}(\mu) \times 130 \times 10^{-11}
       \left(
       \frac{100~\mbox{GeV}}{m_{\scriptstyle \rm susy}}
       \right)^2 \tan \beta,
\label{eq:susy}
\ee
where sgn$(\mu) = \pm$ is the sign of the $\mu$ term in
supersymmetry models and $\tan \beta > 3$--4 is the ratio of
the two scalar vacuum expectation values, $\tan \beta =
\langle \phi_2 \rangle/\langle \phi_1 \rangle$. The $\tan
\beta$ factor is an important source of enhancement. As
experimental constraints on the Higgs mass have increased,
so has the lower bound on $\tan \beta$. With larger $\tan
\beta$ now required, it appears inevitable that
supersymmetric loops have a fairly major effect on the
theoretical prediction of the muon \gmt if $m_{\scriptstyle
  \rm susy}$ is not too large. In fact, equating
(\ref{eq:susy}) and the discrepancy $\Delta a_{\mu}$, for
example the value $\Delta a_{\mu} = +316(79) \times
10^{-11}$ obtained using the
$a_{\mu}^{\mbox{$\scriptscriptstyle{\rm HLO}$}}$
determination of Ref.~\cite{TeubnerPHIPSI09}, one finds
sgn$(\mu)=+$ and
\be
      m_{\scriptstyle \rm susy} \simeq 
      64^{+10}_{-7} \sqrt{\tan \beta} ~\mbox{GeV}.
\ee
For $\tan \beta \sim 4$--50, these values are in keeping
with mainstream supersymmetric expectations. Several
alternative ``new physics'' explanations have also been
suggested~\cite{Czarnecki:2001pv}.

\section{Conclusions}

We examined a number of hypothetical errors in the {\small
  SM} prediction of the muon \gmt that could be responsible
for the present 3--4$\sigma$ discrepancy $\Delta a_{\mu}$
with the experimental value. None of them looks likely.  In
particular, updating Ref.~\cite{Passera:2008jk} we showed
how an increase $\Delta \sigma(s)\!=\!\epsilon \sigma(s)$ of
the hadroproduction cross section in low-energy $e^+e^-$
collisions could bridge $\Delta a_{\mu}$. However, such
increases lead to reduced $\mh$ upper bounds -- even lower
than 114.4~GeV (the {\small LEP} lower bound) if they occur
in energy regions centered above $\sim 1.2$~GeV). Moreover,
their amounts are generally very large when compared with
the quoted experimental uncertainties, even if the latter
were significantly underestimated. The possibility to bridge
the muon \gmt discrepancy with shifts of the hadronic cross
section therefore appears to be unlikely. If, nonetheless,
this turns out to be the solution, then the 95\% {\small CL}
upper bound $\mhUB$ drops to about 135~GeV.

If $\tau$-decay data are used instead of $e^+ e^-$ ones in
the calculation of $a_{\mu}^{\mysmall \rm SM}$, the muon
\gmt discrepancy decreases to $\sim \!2 \sigma$. While this
reduces $\Delta a_{\mu}$, it raises the value of $\dafive$
leading to $\mhUB=$138~GeV, thus increasing the tension with
the {\small LEP} lower bound and suggesting a near conflict
with it should one try to overcome the full discrepancy. One
could also consider a scenario, suggested by recent studies,
where the $\tau$ data confirm the $e^+e^-$ ones below $\sim
\! 1$~GeV, while a discrepancy between them persists at
higher energies. If, in this case, $\Delta a_{\mu}$ is fixed
by hypothetical errors in the $e^+ e^-$ measurements above
$\sim \! 1$~GeV, where the data sets disagree, one also
finds values of $\mhUB$ inconsistent with the {\small LEP}
lower bound.

If the $\Delta a_{\mu}$ discrepancy is real, it points to
``new physics'', like low-energy supersymmetry where $\Delta
a_{\mu}$ is reconciled by the additional contributions of
supersymmetric partners and one expects $\mh \! \lesssim \!
135$~GeV for the mass of the lightest scalar~\cite{DHHSW}.
If, instead, the deviation is caused by an incorrect
leading-order hadronic contribution, it leads to reduced
$\mhUB$ values. This reduction, together with the {\small
  LEP} lower bound, leaves a narrow window for the mass of
this fundamental particle. Interestingly, it also raises the
tension with the $\mh$ lower bound derived in the {\small
  SM} from the vacuum stability requirement.


~

\noindent {\bf Acknowledgments}: We thank D.~Nomura and
T.~Teubner for discussions and communications, and the
organizers of this workshop for the very pleasant and
stimulating atmosphere.
This work was supported in part by {\small U.S.\ DOE} grant
{\small DE-AC02-76CH00016}, {\small E.C.}  contract {\small
  MRTN-CT 2006-035505}, and {\small U.S.\ NSF} grant {\small
  PHY-0758032}.

\end{multicols}
\vspace{-2mm}
\centerline{\rule{80mm}{0.1pt}}
\vspace{2mm}
\begin{multicols}{2}

\end{multicols}
\end{document}